\documentclass[twocolumn,twoside,preprintnumbers,amsmath,amssymb,pacs]{revtex4}
\usepackage{epsfig}
\usepackage{graphicx}% Include figure files
\usepackage{dcolumn}% Align table columns on decimal point
\usepackage{bm}% bold math

\usepackage{fancyhdr}

\usepackage{pslatex}

\newcommand*{\no}{\noindent}
\newcommand*{\bea}{\begin{eqnarray}}
\newcommand*{\eea}{\end{eqnarray}}
\newcommand*{\be}{\bea}
\newcommand*{\ee}{\eea}
\newcommand*{\pd}{\partial}

\newcommand*{\pref}[1]{(\ref{#1})}

\newcommand*{\nn}{\nonumber}

\begin{document}

\title{{\Large On the Faddeev-Popov operator eigenspectrum in topological background fields}}

\author{Axel Maas}

\affiliation{Instituto de F\'\i sica de S\~ao Carlos, Universidade de S\~ao Paulo, C.P. 369, 13560-970 S\~ao Carlos, SP, Brazil}

\received{on}

\begin{abstract}

During the last years significant progress has been made in the understanding of the confinement of quarks and gluons. However, this progress has been made in two directions, which are at first sight very different. On the one hand, topological configurations seem to play an important role in the formation of the static quark-anti-quark potential. On the other hand, when studying Green's functions, the Faddeev-Popov operator seems to be of importance, especially its spectrum near zero.

To investigate whether a connection between both aspects exist, the eigenspectrum of the Faddeev-Popov operator in an instanton and a center-vortex background field are determined analytically in the continuum. It is found that both configurations give rise to additional zero-modes. This agrees with corresponding studies of vortices in lattice gauge theory. In the vortex case also one necessary condition for the confinement of color is fulfilled. Some possible consequences of the results will be discussed, and also a few remarks on monopoles will be given.


\end{abstract}

\maketitle

\thispagestyle{fancy}
\setcounter{page}{0}

\section{Introduction}

The confinement of quarks and gluons in QCD is a long-standing and challenging problem. And although there are very few doubts today that QCD is the correct theory of strong interactions, due to overwhelming evidence from experiments and with increasing importance high precision lattice gauge theory calculations, the lack of a complete understanding of the confinement mechanism is unsatisfactory. Not only that this implies no understanding of one of the most precise measurements in strong interaction physics \cite{Eidelman:2004wy} - the absence of free quarks - this also always casts a doubt on whether the theory is really well-defined.

Thus an understanding of confinement is still a very important task. Especially, as there are more and more hints that the mechanism of confinement is also intimately linked to the mechanism of chiral symmetry breaking, and thus to one of the most important ingredients in hadron phenomenology. In addition, even today no proof exists, whether the presence, not the possibility, of confinement is a necessary consequence of the structure of Yang-Mills theory, or whether it is a dynamical phenomenon.

Nonetheless, vast progress has been made over the decades in the understanding of the non-perturbative features of QCD. Concerning confinement, this has been achieved along two main avenues in recent years.

On the one hand, there is significant evidence that topological field configurations dominate the confining, long-range part of the (static) quark-anti-quark potential \cite{Greensite:2003bk}. In addition, these objects can carry topological charge, which permits them to provide chiral symmetry breaking via the Atiyah-Singer index theorem \cite{Atiyah:1968mp} and the Banks-Casher relation \cite{Banks:1979yr}. Thus these configurations would be a very convenient possibility to understand different non-perturbative phenomena in QCD within one frame-set. However, many different types of topological configurations are known, and it is not yet clear, how they are related, and whether their relevance to non-perturbative phenomena is gauge-dependent. In addition, there seem to exist highly non-trivial relations between the various types \cite{Greensite:2003bk,Reinhardt:2001kf,deForcrand:2000pg}.

On the other hand, the Green's functions of the theory have been investigated, prominently in a class of gauges which includes the Landau and Coulomb gauge \cite{Alkofer:2000wg}. Especially in Landau gauge it has been been empirically found that the gluon propagator cannot be described as an observable particle, according to the Oehme-Zimmermann super-convergence relation \cite{Oehme:bj}: The gluon does not possess a positive semi-definite spectral function. This has been found in calculations using Dyson-Schwinger equations \cite{Alkofer:2000wg,vonSmekal:1997is,Alkofer:2003jj} and the renormalization group \cite{Pawlowski:2003hq}. As this statement is based on the vanishing of the gluon propagator at zero momentum, lattice calculations notoriously have problems to verify this result. However, investigations on highly asymmetric lattices \cite{Silva:2005hb} and in three dimensions \cite{Cucchieri:2003di} are in favor of such a result. The violation of positivity has on the lattice also been observed in direct determinations of the spectral functions \cite{Cucchieri:2004mf,Langfeld:2001cz}. Similar results have also been obtained in Coulomb gauge using variational methods \cite{Reinhardt:2004mm} and in interpolating gauges \cite{Fischer:2005qe}.

This result has also been predicted by the confinement scenarios of Kugo and Ojima \cite{Kugo:gm} and of Gribov and Zwanziger \cite{Gribov:1977wm,Zwanziger:2003cf}. Furthermore, these scenarios predict as the origin of such an infrared vanishing gluon propagator an infrared enhanced Faddeev-Popov ghost propagator in Landau gauge. As in Landau gauge the ghost propagator is just the expectation value of the inverse of the Faddeev-Popov operator, this implies an enhancement of its zero or near-zero spectrum. The prediction of an infrared diverging ghost propagator has been confirmed in functional methods \cite{Alkofer:2000wg,vonSmekal:1997is,Pawlowski:2003hq}, in lattice gauge theory \cite{Bloch:2003sk}, and on quite general grounds \cite{Watson:2001yv}. The enhancement of the spectrum has also been directly observed in lattice gauge theory \cite{Sternbeck:2005vs}. Similar results do also hold in Coulomb gauge \cite{Reinhardt:2004mm,Greensite:2004ur}. Thus the Faddeev-Popov operator plays a central role in these confinement scenarios.

Digressing, it is an interesting side-remark that the Kugo-Ojima scenario is based on arguments using BRST-symmetry while the Gribov-Zwanziger scenario uses the properties of the field configuration space. Thus there seems to be no apparent connection between both approaches, and still their predictions in the investigated cases coincide. It is thus a non-trivial question, if this is a mere coincidence, and one of the scenarios is incorrect, or, if both are correct, how they are linked.

Considering the results listed above, it is thus an highly interesting question, if these two aspects of confinement - topological configurations and the observation of an enhanced spectrum of the Faddeev-Popov operator - are linked. Investigations in lattice gauge theory support such a linkage \cite{Greensite:2004ur}, at least for thin center vortices. Removing these vortices from an ensemble yields a non-confining static quark-anti-quark potential \cite{Greensite:2003bk,Faber:1999sq,deForcrand:1999ms} and restoration of chiral symmetry \cite{deForcrand:1999ms,Gattnar:2004gx}. At the same time the enhancement of the Faddeev-Popov operator vanishes, and the spectrum resembles that of an only weakly perturbed vacuum \cite{Greensite:2004ur}. Consequently, also the enhancement of the ghost propagator vanishes \cite{Langfeld:2002dd}.

Still, a complementary continuum treatment is desirable for several reasons. First, the thin center vortices of lattice gauge theory only provide a fit to a configuration, and may also include contributions from other sources like monopoles. It is thus complicated to disentangle the various contributions. Second, as this is an infrared or small eigenvalue problem, finite volume effects may play an important role, especially as it seems to be very hard to reach the infrared regime in the volumes studied so far in this context. Third, it is only possible to extract numerically a few hundred of the lowest eigenvalues, but the complete spectrum would be interesting. And finally the lattice always views already an ensemble of configurations, making it hard to track the contributions of a single specific field configuration. In general a continuum treatment is a complicated problem, but in some cases it can be solved. Two of these cases will be treated here explicitly: The one-instanton and the one-center vortex configuration \cite{Maas:2005qt}.

\section{Eigenspectrum of the Faddeev-Popov operator}

\subsection{General remarks}

The problem thus to be solved is the eigenvalue problem of the Faddeev-Popov operator (in Euclidean space)
\be
M^{ab}=-\pd_\mu(\pd_\mu\delta^{ab}+gf^{abc}A_\mu^c)\nn,
\ee
\no where $g$ is the gauge coupling, $f^{abc}$ are the structure constants of the gauge group, and $A_\mu^c$ is the field configuration. Hence this task amounts to the determination of the eigenvalues $\omega^2$ and eigenfunctions $\phi^a(x,\omega^2)$ of the eigenequation
\be
M^{ab}\phi^b=-\pd_\mu(\pd_\mu\delta^{ab}+gf^{abc}A_\mu^c)\phi^b=\omega^2\phi^a\nn
\ee
\no Here only the case of the gauge group SU(2) will be treated.

In the vacuum, $A_\mu^c=0$, this operator coincides with the negative Laplacian. Its eigenspectrum is thus the set of positive real numbers and its eigenvectors are plane waves. Furthermore there are three ($N_c^2-1=3$) trivial, i.\ e.\ constant, eigenmodes to eigenvalue zero. Thus an enhancement at zero requires additional zero-modes beyond these trivial ones. 

This eigenvalue problem can be seen as an analogy to quantum mechanics. It is formally equivalent to a stationary Schr\"odinger equation in four space dimensions, and with color being an internal degree of freedom. Then the vacuum case becomes the free-particle problem. Thus the fact that the eigenmodes in the vacuum, and in general eigenmodes at zero or positive eigenvalue, are not normalizable corresponds to the quantum mechanical case of scattering states.

It should be noted that to include in path integrals only one gauge copy of each gauge field configuration it is necessary to restrict the configuration space to a region known as the fundamental modular region. Inside this region the Faddeev-Popov operator is positive semi-definite. Hence in general the eigenfunctions will not be normalizable.

\subsection{Topological field configurations}

For the analysis of the spectrum in topological field configurations, two very distinct cases will be used. Due to their prominent role in the center-dominance scenario of confinement \cite{Greensite:2003bk}, and the studies performed on the same topic using lattice gauge theory \cite{Greensite:2004ur}, thick center vortices are one natural candidate. These objects are sweeping out world-sheets and are thus string-like objects. Their field configuration is \cite{Diakonov:1999gg}
\be
A_\eta^a=\delta^{3a}\frac{1}{g}\frac{\mu(\rho)}{\rho}.\label{vortex}
\ee
\no which is given in two-dimensional cylindrical coordinates $(\rho,\eta)$, i.\ e.\ the field configuration is independent of the $z$- and $t$-coordinate. The function $\mu(\rho)$, the ``vortex profile'', is a smooth function which vanishes at the origin. For $\rho\to\infty$, $\mu$ goes to the constant value $2n+1$, where $n$ is zero or a positive integer. The value $2n+1$ is the flux of the vortex, which has to be odd to allow Wilson-loops pierced by the vortex to take a non-trivial (center-)value. The function $\mu$ thus has also an implicit scale, which defines the size of the vortex. A specific example of a profile will be given below, but the spectrum depends only on the general features at small and long distances \cite{Maas:2005qt}. Note that the vortex has only a component along the abelian color direction.

As an (in some sense maximal) alternative also the case of a point-like instanton will be discussed with field configuration \cite{Bohm:2001yx}
\be
A_\mu^a=\frac{1}{g}\frac{2}{r^2+\lambda^2}r_\nu \zeta^a_{\nu\mu},\nn
\ee
\no where the $\zeta^a$ are the 't Hooft tensors and $\lambda$ is the size of the instanton. Although instantons seem to play a significant role in the physics of chiral symmetry breaking \cite{Schafer:1996wv}, it is unlikely that they contribute to the mechanism of confinement (see, however, e.\ g.\ \cite{Negele:2004hs} on this topic). Thus it very interesting to check whether they have any relevance for the spectrum of the Faddeev-Popov operator and thus e.\ g.\ in the Gribov-Zwanziger scenario.

With these field configurations the problem of determining the spectrum of the Faddeev-Popov operator is fully specified. Especially there is no necessity to fix a specific gauge. In fact, the results found are valid in any gauge in which the field configurations satisfy the gauge condition. E.\ g.\ both are admissible configurations in Landau gauge, as they are transverse. The vortex is in addition also admissible in Coulomb gauge, gauges interpolating between the Coulomb and Landau gauge, and some axial gauges. In both cases, further gauges can be constructed, which permit such field configurations. Thus the results apply to a large number of possible gauges.

However, note that in principle a fully specified gauge, i.\ e.\ free of Gribov ambiguities, is necessary in the non-perturbative regime treated here. Thus the question which field configurations are admissible in a given gauge is in general much more subtle. The instanton e.\ g.\ is still admissible in a fully fixed Landau gauge, but this is not as clear for the vortex, although it is likely \cite{Maas:2005qt}.

On the other hand, it is possible to perform a gauge transformation on the given field configurations. As the Faddeev-Popov operator is not gauge invariant, this can change its spectrum. In the only case where this has been checked explicitly so far, with the instanton in a singular field configuration rather than the regular field configuration presented here, this did not change the spectrum, although the eigenfunctions were modified \cite{Maas:2005qt}. It is thus a very interesting question whether this is accidental or whether at least some part of the spectrum of the Faddeev-Popov operator is gauge-invariant. This is an open problem.

\subsection{Solutions to the eigen-equations}

As both field configurations are transverse, the eigen-problem simplifies considerably to the equation
\be
((\pd^2+\omega^2)\delta^{ab}+gf^{abc}A_\mu^c\pd_\mu)\phi^b=0\nn.
\ee
\no The detailed solution of the eigen-problem is somewhat technical and lengthy and can be found elsewhere \cite{Maas:2005qt}. There are, however, a few noteworthy features.

Concentrating first on the instanton case, it is found that the angular and radial part can be separated. The angular part and the color dependence can be projected onto a three-dimensional spin problem. The eigenfunctions are thus appropriate combinations of (hyper-)spherical harmonics, and can be labeled by a 'spin quantum number' $l$ and a 'magnetic quantum number' $m$. The different colors differ only in their composition in terms of the eigenstates of the magnetic quantum number for a given, common $l$. The remaining radial equation, which is the same for each color, is then given by
\be
0=\frac{1}{r^3}\pd_r r^3\pd_r\phi_{lc}+\left(\omega^2-\frac{4l(l+1)}{r^2}+\frac{c}{r^2+\lambda^2}\right)\phi_{lc},\label{req}
\ee
\no where $c$ is a constant emerging from the separation of the angular part, which takes the values $4$, $-4l$, and $4(l+1)$ with multiplicities $2l+1$, $2l+3$, and $2l-1$, respectively. The differential equation is solved by
\bea
\phi_{lc}&=&r^{2l}\sum_{n=0}^{\infty}a_n\left(-\frac{r^2}{\lambda^2}\right)^n\nn\\
a_{-1}&=&0\nn\\
a_{0}&=&D\nn\\
a_{n}&=&\frac{\omega^2\lambda^2(a_{n-1}-a_{n-2})+((4n+8l)(n-1)+c)a_{n-1}}{4n(n+1)+8ln}\nn.
\eea
\no Here, $D$ is an arbitrary overall normalization constant. A second solution diverges at zero and is thus not admissible for a smooth gauge field. Unfortunately this is a series of hyper-geometric type, and thus its convergence for $r\to\infty$ cannot be tested easily.

\begin{figure}[t]
\epsfig{file=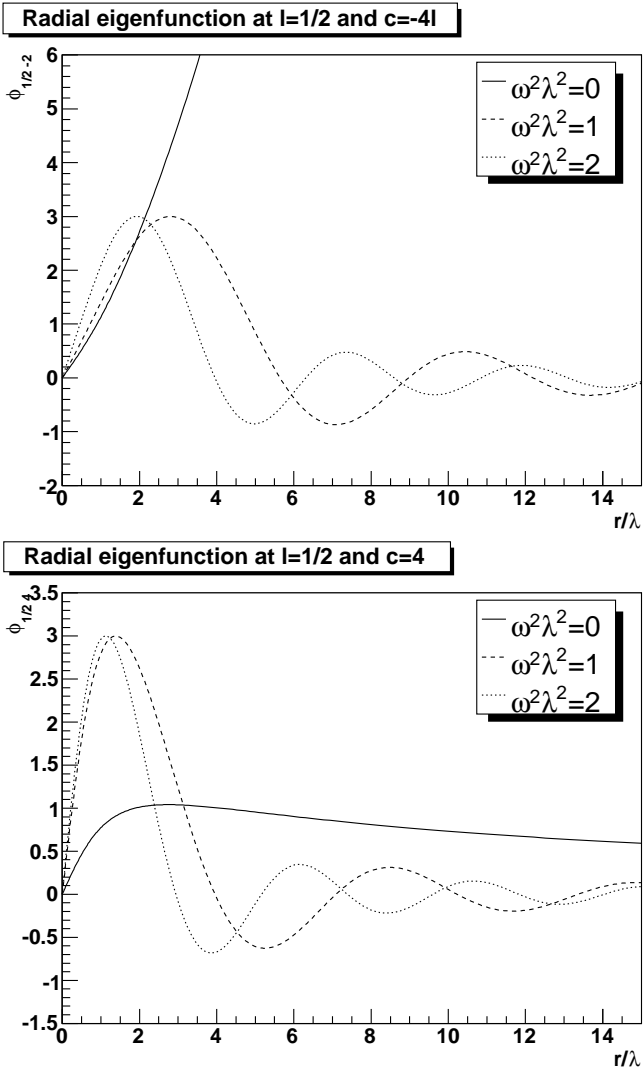,width=\linewidth}
\caption{The radial eigenfunctions $\phi_{lc}$ for $l=1/2$ and the two different $c$ values. For better visualization, positive $\omega^2$-solutions have been normalized so that their maximum is $3$, while modes with $\omega^2\lambda^2=0$ have been normalized so that $\phi_{lc}/r^{2l}|_{r=0}=1$. An example for a diverging, and thus non-admissible, zero-mode is plotted in the top panel.}\label{figl0.5}
\end{figure}

However, in the case of $\omega^2=0$, the sum can be summed explicitly, and only for $l=1$ and $c=8$ with multiplicity one and $l=1/2$ and $c=4$ with multiplicity two non-diverging solutions are found. In these cases the zero-modes can be given in closed form and read for $l=1/2$
\be
\phi_{\frac{1}{2}\;4}(r)=2D\frac{-\frac{r^2}{\lambda^2}+\left(1+\frac{r^2}{\lambda^2}\right)\ln\left(1+\frac{r^2}{\lambda^2}\right)}{\frac{r^3}{\lambda^3}}\nn
\ee
\no and for $l=1$
\be
\phi_{1\;8}(r)=3D\frac{\frac{r^4}{\lambda^4}+2\frac{r^2}{\lambda^2}-2(1+\frac{r^2}{\lambda^2})\ln\left(1+\frac{r^2}{\lambda^2}\right)}{\frac{r^4}{\lambda^4}}.\nn
\ee
\no No further zero-modes exist for larger $l$ or for any other eigenvalue $c$. Hence, the instanton sustains six zero-modes in total, the three trivial ones at $l=0$, two for $l=1/2$ and one for $l=1$. To yield an impression of these solutions, the radial part of the eigenfunctions at $l=1/2$ are shown in figure \ref{figl0.5}

For non-zero eigenvalue, only a numerical test is possible. It is found that for $\omega^2>0$ the eigenfunctions are always finite and behave essentially like a Bessel-function at large $r$. For negative $\omega^2$ the eigenfunctions always diverge as $r\to\infty$, and are thus not admissible. Thus for each value of $\omega^2$ greater zero a denumerable infinite number of eigenmodes exist: All $l$ are possible for all non-zero $\omega^2$. Thus the only enhancement compared to the vacuum occurs at zero eigenvalue.

The situation is similar in the case of the vortex. First of all, the equation for the abelian or color 3-component completely decouples and is solved by the vacuum solution. This is due to the explicit factor $\delta^{3a}$ in the vortex field configuration \pref{vortex} and the antisymmetry of the structure constants. In the other equations it is possible to separate the trivial coordinates $z$ and $t$ and also the angular part in the remaining two-dimensional sub-space, introducing an 'angular quantum number' $m$ (there is no magnetic quantum number in two dimensions). After appropriate adding and subtracting real and imaginary parts of the equations, it is again sufficient to solve a single ordinary differential equation in radial direction, given by
\be
\left(\frac{1}{\rho}\pd_\rho\rho\pd_\rho-\frac{m^2}{\rho^2}+\omega^2(1-s^2)\right)b_m^+-\frac{m\mu}{\rho^2}b_m^+=0,\label{origvort}
\ee
\no where $b_m^+$ is a linear combination of the two color functions $\phi^1$ and $\phi^2$ and their complex conjugates. This equation can be solved again, yielding as the only potentially non-divergent solution
\bea
b_m^+&=&\rho^{|m|}\sum_{n=0}^{\infty}b_{mn}^+\rho^n\nn\\
b_{m-1}^+&=&0\nn\\
b_{m0}^+&=&D\nn\\
b_{mn}^+&=&\frac{-\omega^2(1-s^2)b_{mn-2}^++m\sum_{i=0}^{n-1}b_{mn-1-i}^+\mu_{1+i}}{n^2+2n|m|}\nn.
\eea
\no $D$ is again a normalization constant. Herein the $\mu_i$ are the series coefficients of the function $\mu$. Thus this solution is only correct, if such a series expansion is possible. Otherwise the system can still be solved numerically, and does not yield a qualitatively different behavior than the one discussed below.

Unfortunately, this series cannot be summed even for the zero-modes. Nonetheless, an analysis of the limit $\rho\to\infty$ of the original equation \pref{origvort} allows to judge the existence of zero-modes. It turns out that there are $4n$ for a vortex of flux $2n+1$ \cite{Maas:2005qt}. This implies that a flux 1 vortex does not support any zero-modes at all, and an arbitrary number of zero-modes can be generated by a vortex of sufficient flux. This is thus quite distinct from the instanton case.

\begin{figure}[t]
\epsfig{file=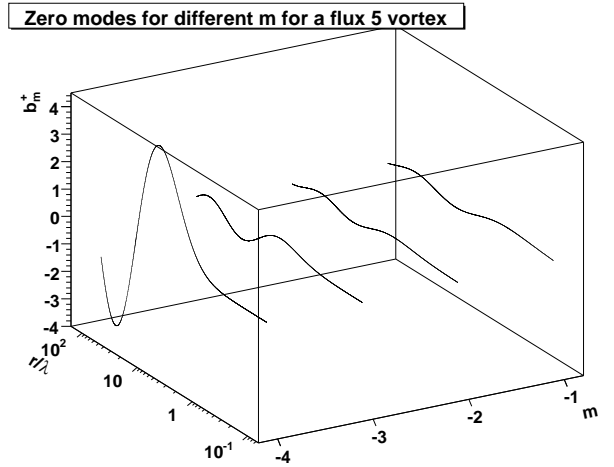,width=\linewidth}
\caption{The radial behavior of the zero-modes $b_m^+$ for different $m$ in a flux 5 vortex. The solutions have been normalized so that $b_m^+/\rho^{|m|}|_{\rho=0}=1$. The same set of solution also exists at the corresponding positive values of $m$.}\label{figvortex}
\end{figure}

Furthermore, again only at positive $\omega^2$ further solutions are admissible, but again a denumerable infinite number of them for each value of $\omega^2$ exist, labeled by the integer $m$ and particular combinations of the colors. Thus the only enhancement is again at eigenvalue zero. To get an impression of the very different radial behavior of the zero-modes than in the instanton case, these are plotted for a flux 5 vortex in figure \ref{figvortex} for the profile $5\rho/(\rho+\lambda)$. For the function $b_m^+$ only solution at negative $m$ exist. Another composition provides positive $m$-solutions. The original functions $\phi^1$ and $\phi^2$ are formed from superpositions of these functions.

In both cases, the instanton and the vortex, the non-zero modes behave at large distances essentially like Bessel functions, and thus like the partial waves in the vacuum case. As expected, they are non-localized. The zero-modes, on the contrary, show a very distinct behavior in both cases. While for the instanton they vanish slowly, or become constant \cite{Maas:2005qt}, they oscillate on a logarithmic scale for vortices. Thus they are non-normalizable, as expected. This non-normalizability of zero-modes is also an important feature in the Gribov-Zwanziger scenario \cite{Zwanziger:2003cf}, but its significance is still not completely understood, although in the present context its emerges in a natural way.

\subsection{A criterion for confinement}

Hence both of the topological field configurations do provide additional zero-modes to the Faddeev-Popov operator, and can thus be relevant in the Gribov-Zwanziger and/or Kugo-Ojima scenario. In fact, it can also be shown that the instanton is located in the region of field configuration space, which is the relevant one in the Gribov-Zwanziger scenario \cite{Maas:2005qt}. This is not yet known for the vortex configuration, although it appears likely. In addition, the large enhancement at zero eigenvalue for high-flux vortices makes them an important candidate for the relevant field configurations in the Gribov-Zwanziger scenario.

Apart from this aspect, it is a sensible questions, whether the particular single-vortex field configuration can also contribute to the static quark-anti-quark potential. It is not (yet) possible to give a final answer to it, but the following observation can be made: The vortex is admissible in Coulomb gauge. In this gauge, it is possible to formulate a necessary condition for confinement, namely that free color charges have an infinite energy, in terms of the eigensystem of the Faddeev-Popov operator. The criterion is given by \cite{Greensite:2004ur} 
\be
\lim_{\omega^2\to 0}\frac{\gamma(\omega^2)F(\omega^2)}{\omega^2}>0\label{cond}
\ee
\no where $\gamma$ is the eigenvalue density of the Faddeev-Popov operator, and $F$ is the expectation value of the negative Laplacian in the eigenmodes of the Faddeev-Popov operator. Of course, for a vortex with flux greater than one, $\gamma(0)$ is non-vanishing. Furthermore, it is found that in this case $F(\omega^2)=\omega^2$ \cite{Maas:2005qt}, and thus the criterion is met.

\section{Outlook and summary}

These results indicate that also in the continuum there is an intimate link between topological configurations, the properties of the Faddeev-Popov operator, and the confinement of color charges. However, the consequences of the quite distinct properties of instantons and vortices, and what this implies for the confinement mechanism, are not yet clear. It can be argued, that the small number of additional zero-modes provided by the instanton indicate that this configuration may be less relevant than large flux vortices. Still, this is not completely satisfactory, and more understanding is required.

Furthermore, these two types of topological configurations are not the only ones relevant in the continuum. Especially monopoles have received much attention, due to the phenomenologically attractive dual superconductor model \cite{Ripka:2003vv}. In various investigations it has been found that monopoles, vortices, and instantons are closely related \cite{Reinhardt:2001kf,deForcrand:2000pg}. It is still unclear, whether it is possible to disentangle their contributions, and single out one object which is responsible for all the facets of confinement, even in one single gauge. It even seems more likely that their non-trivial interactions may be an important ingredient in the confinement problem.

Thus it would be very interesting to investigate the spectrum of the Faddeev-Popov operator in a monopole background. Unfortunately, even in abelian projection, this turns out to be more complicated than the cases presented here, and will require further work. Nonetheless, there are a few general features, which already offer an interesting perspective.

Assuming that the spectrum of the Faddeev-Popov operator is not altered in a qualitative way by abelian projection, the monopole field configuration will be proportional to the abelian generator. This is the same situation as for the vortex treated here. Thus as in the case of the vortex, the abelian component of the Faddeev-Popov operator (and thus the corresponding ghost) decouples and becomes trivial. On the other hand, provided that monopoles give rise to non-trivial zero-modes in the Faddeev-Popov operator, the off-diagonal ghosts would experience an infrared enhancement. As the off-diagonal ghosts are coupled to the diagonal gluon in maximal abelian gauge \cite{Dudal:2004rx}, it is tempting to conjecture that diagonal gluons will be confined by a Gribov-Zwanziger mechanism. This would cure one of the greatest sicknesses of the dual color superconductor models, the presence of an unconfined, massless gluon. The off-diagonal gluon would still be confined by a Higgs-like mechanism. However, this is a highly speculative chain of arguments, but first explorative lattice studies \cite{Bornyakov:2003ee} as well as investigations using effective potentials \cite{Capri:2005tj} seem to support such a scenario. Still much work needs to be done, but this offers a very attractive perspective.

Nonetheless, it is still unlikely that monopoles alone can be the end of the story, as their close relation to center vortices and their inability to account e.\ g.\ for Casimir scaling necessitates interaction with other topological degrees of freedom \cite{Greensite:2003bk,deForcrand:2000pg}.

Summarizing, a relation between the topological configurations, an enhancement of the Faddeev-Popov operator, and the confinement of color has been explicitly demonstrated using analytical techniques in the continuum. Combined with results from lattice calculations, this explicitly demonstrates that there is a close relationship between these aspects of confinement. And although many open questions remains, different aspects of confinement seem to have at least a connection, if not a common origin. Thus a further piece of the confinement puzzle is slowly uncovered. And with each further piece, it is possible to see more of the whole picture, and what has been uncovered so far unfolds a very rich structure, indicating that the dynamical realization of confinement is a process which interconnects various elements in a very interesting way. Thus more and more the unification of the various aspects of confinement becomes an even more important goal, as only a clear and full understanding will permit to understand the full non-perturbative structure of QCD.\\[3mm]

\no{\bf Acknowledgment}\\[2mm]

The author thanks the organizer for this interesting meeting and the opportunity to present this work and for financial support. Support from Funda\c{c}\~ao de Amparo \`a Pesquisa do Estado de S\~ao Paulo (FAPESP) and also of the Deutsche Forschungsgemeinschaft (DFG) under grant  MA 3935/1-1 are also acknowledged. Furthermore the author is grateful to R.~Alkofer and J.~M.~Pawlowski for fruitful discussions.

\end{document}